# Visualizing Strain-induced Pseudo magnetic Fields in Graphene through an hBN Magnifying Glass

Yuhang Jiang[1], Jinhai Mao[1], Junxi Duan[1], Xinyuan Lai[1], Kenji Watanabe[2], Takashi Taniguchi[2] and Eva Y. Andrei[1]

[1]Department of Physics and Astronomy, Rutgers University, Piscataway, New Jersey, 08854, USA

[2]Advanced Materials Laboratory, National Institute for Materials Science, 1-1 Namiki,, Tsukuba 305-0044, Japan

## Abstract

Graphene's remarkable properties are inherent to its 2D honeycomb lattice structure. Its low dimensionality, which makes it possible to rearrange the atoms by applying an external force, offers the intriguing prospect of mechanically controlling the electronic properties. In the presence of strain, graphene develops a pseudo-magnetic field (PMF) which reconstructs the band structure into pseudo Landau levels (PLLs). However, a feasible route to realizing, characterizing and controlling PMFs is still lacking. Here we report on a method to generate and characterize PMFs in a graphene membrane supported on nano-pillars. A direct measure of the local strain is achieved by using the magnifying effect of the Moiré pattern formed against a hexagonal Boron Nitride (hBN) substrate under scanning tunneling microscopy (STM). We quantify the strain induced PMF through the PLLs spectra observed in scanning tunneling spectroscopy (STS). This work provides a pathway to strain induced engineering and electro-mechanical graphene based devices.

Graphene, with its 2D arrangement of atoms in a honeycomb lattice and its suite of unique electronic and mechanical properties[1,2], carries the promise of realizing flexible, stretchable and transparent electronics[3] that could lead to many potential applications. In particular, its $sp^2$ bonded carbon atoms can sustain a record high 25% elastic distortion, making graphene the strongest material known[4]. The rearrangement of graphene's atoms in response to strain modifies its low energy band structure and can lead to extremely large strain-induced PMFs[5-8]. However, the lack of a controllable method to introduce strain[8] has hindered progress in this area.

A promising configuration for introducing strain in graphene by supporting it on nano-pillars[9-13] was recently proposed and studied with spatially resolved Raman spectroscopy, atomic force microscopy (AFM) and scanning electron microscopy (SEM)[9-11]. These experiments showed that the strain could be varied by tuning the pillar configuration, their height and density. But, although the strain was found to generate new features in the Raman spectra, no evidence of spectral reconstruction was reported even for strains up to 20% [9]. This raises the question whether such geometry is suitable for manipulating graphene's band structure. Here we revisit this issue by combing SEM, AFM, STM and STS to investigate strain-induced PMFs in a graphene membrane stretched over a nano-pillar array. STM measurements of the distorted Moiré pattern produced by strained graphene resting on an hBN substrate in between pillars, provided a direct measure of the local strain. The magnifying effect[14] of the Moiré pattern makes it possible to detect strain levels that are otherwise below the instrumental resolution. At the same time the appearance of clear PLLs in the strained regions afford direct and quantitative evidence of the strain-induced PMF.

Figure 1a illustrates the sample fabrication steps (details in SI) designed to induce a strain network in the graphene membrane covering the nano-pillars [12,13]. Two types of pillar materials were used: an insulator, LOR (photoresist), and a conductor, Au. Following the fabrication of the pillar array, a single layer of graphene is deposited on the pillars with a thin PMMA sacrificial film using the standard dry transfer method[15]. The PMMA film provides rigidity to the graphene/PMMA structure so that it remains suspended even after being transferred onto the pillar array. Subsequently, the removal of the PMMA film with solvent produces a strong capillary force between the graphene membrane and the substrate[16]. The final result is a distorted 3D graphene lattice with conical singularities at the pillar positions. Stress radiates outward from the pillars producing strain which results in a graphene wrinkle network (Fig.1b), similar to the well-known wrinkling in thin elastic membranes[17-20]. Depending on the aspect ratio of the pillar array, this procedure can result in either a suspended graphene canopy or a collapsed graphene membrane resting on the substrate between pillars[10]. However, in both the suspended and supported configurations the graphene membrane is subjected to a non-uniform distribution of strain which is expected to produce PMFs [13]. The pillar arrays were triangular with periods in the range 1-2μm and height range of 50-600nm.

Following the fabrication, samples are characterized at room temperature by optical microscopy, SEM (30kV acceleration voltage) and by AFM topography in the semi-contact mode. The samples are then annealed overnight in forming gas (9:1, Ar to $H_2$ ratio) at 230°C in order to remove the PMMA residue prior to the STM/STS measurements at 4.6K. In STS, the dI/dV spectra providing a measure of the local density of states (LDOS) (I is the tunneling current, V is the sample bias) are measured by the standard lock-in technique with an AC voltage modulation ($V_{RMS}$ = 4mV, f = 473.1Hz) added to the DC bias.

We first explore the sample supported on the LOR pillared structure. Figure 2a shows the optical micrograph of graphene on the pre-patterned LOR pillars (600nm height and 1μm period). In this case, the pillars are first patterned using standard SEM lithography by overdosing the LOR, but the pattern is not developed in order to keep a flat surface for the coming graphene transfer (see SI for details). Next, the graphene membrane is stacked on top of the undeveloped LOR substrate by a dry transfer process[21]. Finally, the LOR between the pillars is removed with an ethyl-lactate developer followed by exposure to nitrogen gas flow for drying. The evaporation of the solvent that fills the gap between the graphene and the substrate generates a strong capillary force causing the graphene membrane to collapse towards the substrate.

Figure 2b shows a top view SEM image of the graphene (light green) on LOR pillars (bright spots) after the developer step. We note that the graphene flake was torn in two after the development, providing a vivid illustration of the huge deformation strain generated during the collapse process. The AFM topography image (inset in Figure 2b) shows sagging of the graphene membrane in between pillars. Interestingly we note that the collapse of the membrane introduces a network of strain-induced ripples linking each pillar to its neighbors.

Turning to Figure 2c, we observe that the LOR pillars adjacent to the tear in the graphene flake are visibly bent out of shape as they are pulled apart by the two torn pieces. The bent pillars provide a strikingly visual confirmation of the strain in this system. A 45° angle side view (Figure 2d) directly illustrates the bent pillars at the edge of the flake (blue arrow). In contrast, the pillars outside the graphene covered regime (yellow arrows) remain upright and undistorted (yellow line). The bent pillars shed light on the response of the system to the strain induced by the collapse of the structure. One would expect the strain to be released by the graphene sheet sliding off the pillars. Instead, it appears that graphene is firmly anchored to the pillars via the van der Waals force and that part of its strain is released by bending the pillars. This scenario can only happen at the edges of the flake where the strain is asymmetric. However, in the center where no such release mechanism is available, once the strain exceeds the breaking point the flake must tear as is illustrated in Figure 2b. The bent LOR pillars technique presented here provides a direct and compelling visualization of the strain pattern.

We now turn to STM measurements for characterizing the local strain. Here we used an Au pillar array (70nm height and 2µm period) supported on an hBN flake instead of LOR for better surface cleanliness. The hBN flake was exfoliated on the $SiO_2$ surface and spin-coated with a thin layer of PMMA. Subsequently standard lift-off lithography was used starting with SEM exposure and development followed by Au/Ti (70nm) deposition, and finally PMMA removal with acetone and isopropyl. The pillared substrate was further annealed in forming gas at 230℃ for 3 hours to remove the PMMA residue before stacking the graphene on top. Au electrodes were added at the periphery of the graphene flake to clamp it down and avoid slippage. Figure 3a shows an SEM image of graphene on the Au pillared substrate. To explore the strain effect, we first focus on the STM topography of an unstrained region of graphene lying outside the pillar array (Figure 3b). The small sample area was located by employing a capacitive navigation technique and a guiding electrode pattern. This technique is extremely efficient for finding small samples in the absence of optical access rapidly and without damaging the tip or sample[22]. Outside the pillar array we observe an undistorted Moiré pattern of period 2.7nm which corresponds to a twist angle of 5° between the graphene and hBN lattices[23,24]. The undistorted Moiré pattern signifies the absence of strain. This is further confirmed by the perfectly hexagonal Fourier transform pattern of the image in Figure 3c. The dI/dV spectrum in this region is "V" shaped again consistent with unstrained graphene[25] (Insert of Figure 3b).

We next consider the STM measurements taken within the pillared region. Here we observe a distorted Moiré pattern (Figure 3e) and its Fourier transform (Figure 3f), indicative of a strain-induced graphene lattice distortion. For unstrained graphene stacked on hBN, the slight mismatch between the two honeycomb lattices creates a Moiré super-structure with perfect hexagonal symmetry[23,24] (Figure 3b, 3c). If the graphene lattice is distorted the resulting Moiré pattern loses its hexagonal symmetry and the pattern is significantly altered[26] as is clearly seen in the topography (Figure 3d, 3e) as well as in its Fourier transform (Figure 3f) (SI). A Moiré pattern is very useful in this case because it can serve as a highly effective magnifying glass of strain-induced lattice distortions[14].

To extract the strain from the distorted Moiré pattern, we consider a model where a strained graphene lattice is superposed on an unstrained hBN crystal substrate. In this case the distorted superlattice Bragg vectors can be obtained using an elegant analytic expression[14]:

$\boldsymbol{G}_m = \frac{4\pi(\delta'^2+\theta^2-w'^2)}{\sqrt{3}a^2} \boldsymbol{l}_z \times \boldsymbol{A}_m$, where $\boldsymbol{l} =(l_x, l_x)$ is the principal axis direction of the strain tensor, $\theta \ll 1$ rad is the misalignment angle between the two lattices, $\delta = 1.8\%$ is the lattice mismatch between unstrained graphene and the hBN substrate. The strained lattice mismatch $\delta' = \delta - w(1-\sigma)/2$, depends on the strain magnitude, $w$, and on the Poisson ratio in graphene[27], $\sigma = 0.165$, and $w' = -w(1+\sigma)/2$. Here $\boldsymbol{A}_m = \widehat{M}\boldsymbol{a}_m$ are the lattice vectors of the distorted superlattice, $\widehat{M} = \frac{1}{\delta'^2+\theta^2-w'^2}\begin{pmatrix} \delta' + (l_y^2 - l_x^2)w' & \theta - 2l_xl_yw' \\ -\theta - 2l_xl_yw' & \delta' + (l_y^2 - l_x^2)w' \end{pmatrix}$ defines the transformation which magnifies and distorts the original graphene lattice, and $\boldsymbol{a}_m$ is the set of six lattice vectors of undistorted graphene that are obtained by $m\pi/3$ rotations of $\boldsymbol{a_0} = (a, 0)$ with $a = 0.246$ nm the lattice constant and m=0,1,…5.

Solving for the measured values of $G_m$ in Figure 3f (dashed lines) we obtain $\theta = 0.07\,rad$ and $w = 4.5\%$, resulting in a magnification, M ~ 20, for the largest lattice vector of the distorted lattice (black arrow in Figure 3e).

To verify the validity of the calculation we numerically superpose a distorted graphene lattice on an hBN substrate, where the distortion is calculated using the parameters $\theta = 0.07\,rad$, $w = 4.5\%$ and $\sigma = 0.165$. The resulting distorted super-lattice Moiré pattern, shown in the bottom panel of Fig. 3d, closely resembles the measured one shown in the top panel. Significantly, although such a small distortion would be very difficult to detect in an isolated graphene sheet unless using a state of the art STM machine, the 20-fold magnification afforded by the Moiré pattern makes the distortion readily detectable even with standard STM resolution. Indeed, earlier STM experiments on an MoS$_2$ layer strained by a pillared substrate[16] were not able to discern the lattice distortion or the presence of a PMF even though Raman spectra showed the presence of strain.

Considering the dI/dV spectrum in the distorted region (Figure 4a) we note that it no longer resembles the featureless "V" shape expected in unstrained graphene. Instead, the spectrum consists of a series of peaks suggesting the presence of a PMF which we discuss next. A strain induced PMF can arise when the nearest neighbor hopping parameters are modified. This introduces a pseudo-vector potential term, $\vec{A}$, in the Dirac-Weyl Hamiltonian describing the low energy excitation of graphene[28] (SI). In the limit of small atomic displacements, u<<a: $A_x \sim \phi_0 \frac{\beta}{a}(u_{xx} - u_{yy})$; $A_y \sim \phi_0 \frac{\beta}{a} 2u_{xy}$, where $\phi_0 = \frac{h}{e}$ is the fundamental unit of flux, $u_{ij}(x,y)$ is the 2D strain field with the x axis taken along the zigzag direction of the graphene lattice, and $\beta = -\partial \ln t / \partial \ln a \,|_{a=a_0} \sim 3.4$ relates the change in the hopping amplitude to the bond length[29]. The pseudo-vector potential gives rise to a PMF, $\vec{B} = \vec{\nabla} \times \vec{A}$, normal to the graphene plane. Unlike a real magnetic field, the strain-induced PMF has opposite signs for graphene's two valleys, consistent with the fact that elastic deformations do not violate the time-reversal symmetry of the crystal. The PMF gives rise to a sequence of quantized LLs similar to those produced by an external magnetic field[2]:

$$E_n - E_D = sgn(N)v_F\sqrt{2e\hbar B|N|} \quad N = 0, \pm1, \pm2, ... \quad (1)$$

Here $E_D$ is the Dirac point energy and N is the level index and $v_F$ is the Fermi velocity.

In order to quantify the PMF in the strained graphene sample we label the peak sequence in the STS spectra starting from N=0 which is taken close to the Dirac point in the unstrained lattice as shown in Figure 4a. The linear dependence of the peak energies on $N^{1/2}$ shown in Figure 4b, supports the interpretation of the peaks in terms of PLLs. Fitting the sequence to equation (1) and assuming $v_F = 1.0 \times 10^6$ m/s, we obtain B ~ (7.7 ± 1)T. Plotting the spectra and PMF along a line perpendicular to a fold (Figure 4c, 4d) we find the average PMF in this region, $B_{PMF}$ ~ (6 ± 2)T.

We next compare the value of the PMF obtained from the LL sequence to that expected from the strain-induced lattice distortion measured with the STM topography. An order of magnitude estimate of the strain-induced PMF can be obtained by assuming a triaxial strain configuration[30,31], $B \approx \frac{\hbar}{e}\frac{8\beta\epsilon}{aD}$. Taking the size of the strained area as the disc of diameter D~1000nm, enclosed by the triangle of folds connecting three pillars, and using the strain value w ~ 4.5%, we obtain an estimate of B ~ 3.2T, consistent with the value obtained from the PLL sequence.

Previous reports of strain induced PLLs employed STS measurements on graphene nanobubbles[8,32]. In these reports the strain was tightly localized on the bubble and relaxed outside it. In contrast to the bubble geometry, the in-plane strain configuration realized here makes it possible to delocalize the strain induced PMF so that it is observed at distances that are hundreds of nanometers away from the nanopillar source. Importantly, the pillared device configuration for introducing strain is compatible with standard device fabrication methods, the only requirement being the pre-patterned substrate.

In summary, we have studied strain-effects in a graphene membrane in contact with an hBN substrate and stretched by an array of nano-pillars. The induced strain was directly visualized in STM and quantified through the magnifying effect of the Moiré pattern formed against the hBN substrate. STS measurements revealed a sequence of PLL peaks in the DOS consistent with the strain-induced PMF. This work provides a quantitative comparison between the measured local strain and the induced PMF, and demonstrates the possibility of modifying graphene's electronic band-structure by mechanical means rather than by chemical functionalization. Moreover, it introduces a new pathway for engineering the band structure[12] and for realizing exotic transport properties[7,33] by patterning devices with pre-programmed PMFs.

E.Y.A and J.M acknowledge support from DOE-FG02-99ER45742.

**Figure captions:**

**Figure 1 Schematics of device fabrication. (a)** Stacking graphene on the pillared substrate. **(b)** Schematics of graphene supported on a pillar (red arrow) and the strain-induced ripples (blue arrows).

**Figure 2 SEM images of graphene on LOR pillars.** (**a**) Optical micrograph of graphene transferred on the LOR surface with pre-patterned pillars (details in main text). Dark-gray and green colors indicate multilayer and single layer graphene respectively. (**b**) SEM top view of the sample in (a) with LOR pillars seen as bright spots. Inset: AFM topography of graphene covering the pillars illustrates the sagging of the graphene layer in between pillars as well as the network of ripples formed along the symmetry directions of the pillar array. (**c**) High resolution SEM image of the sample clearly shows the tear caused by the large strain. (**d**) 45% side view shows bending of the LOR pillars supporting the edge of graphene membrane. Yellow dashed line indicates original orientation of the LOR pillars. Numbers designate the pillar rows. Note that pillars not covered by the graphene membrane, rows 1-4, are undistorted (yellow arrows). Pillars in row 5 and above which are situated under the graphene canopy are bent (blue arrow).

**Figure 3 Imaging the distorted Moiré pattern in graphene.** (**a**) SEM image of the graphene/Au pillars (white spots) and the strain-induced ripples (lines connecting the pillars). (**b-c**) STM topography and FFT pattern of flat graphene far from the pillar area. The superlattice corresponds to the Moiré pattern formed between graphene and the hBN flake, $V_b$ = -300mV, I = 20pA. Inset: dI/dV curve on flat graphene in (b) shows the characteristic "V" shape of unstrained graphene. (d) Comparison between measured distorted Moire pattern on strained graphene (top) and simulated model (bottom) as described in the text. The unit cell is shown in both panels. (**e-f**) Distorted

Moiré pattern and its FFT produced by the strained graphene lattice and the hBN substrate at the position marked by the red square (a). Arrows indicate the set of six lattice vectors of distorted ghraphene $\mathbf{A}_m$ is described in the text.

**Figure 4 Electronic structure of strained graphene.** (**a**) dI/dV curve for graphene observed near the ripple (red square in Figure 3a), $V_b$ = -300mV, I = 20pA, $V_g$ = -10V. Peaks are labeled by their corresponding LL index. (**b**) Linear fit to equation (1) of peak energy versus square-root of LL index N from which we obtain the PMF value B = (7.7 ± 1) T. (**c**) Spatial dependence of the dI/dV curves along the direction perpendicular to the ripple. Same parameters as (a). (**d**) PMF values versus distance from the ripple extracted from the dI/dV curves in (c). The green arrow from A to B (23.1 nm) corresponds to the position of the spectra in (c).

**Figures:**

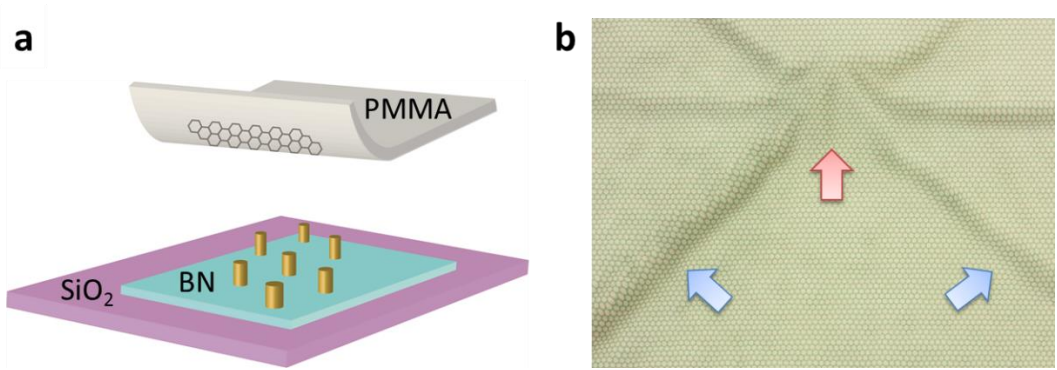

# Figure 1

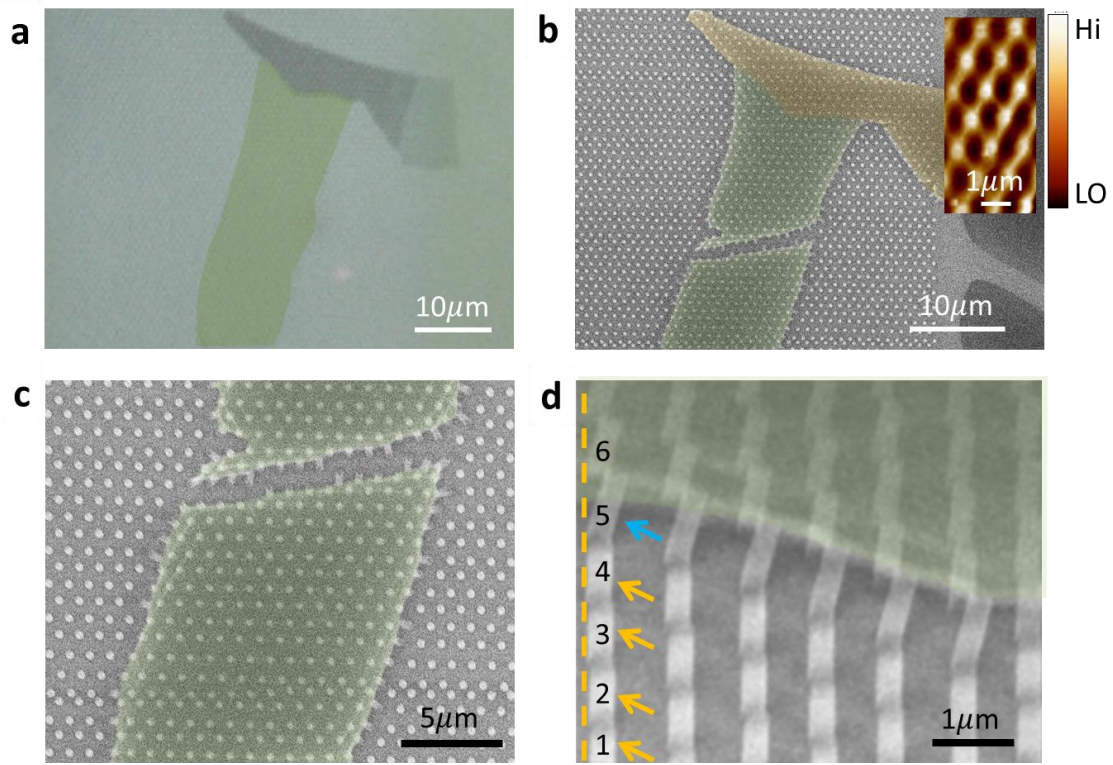

**Figure 2**

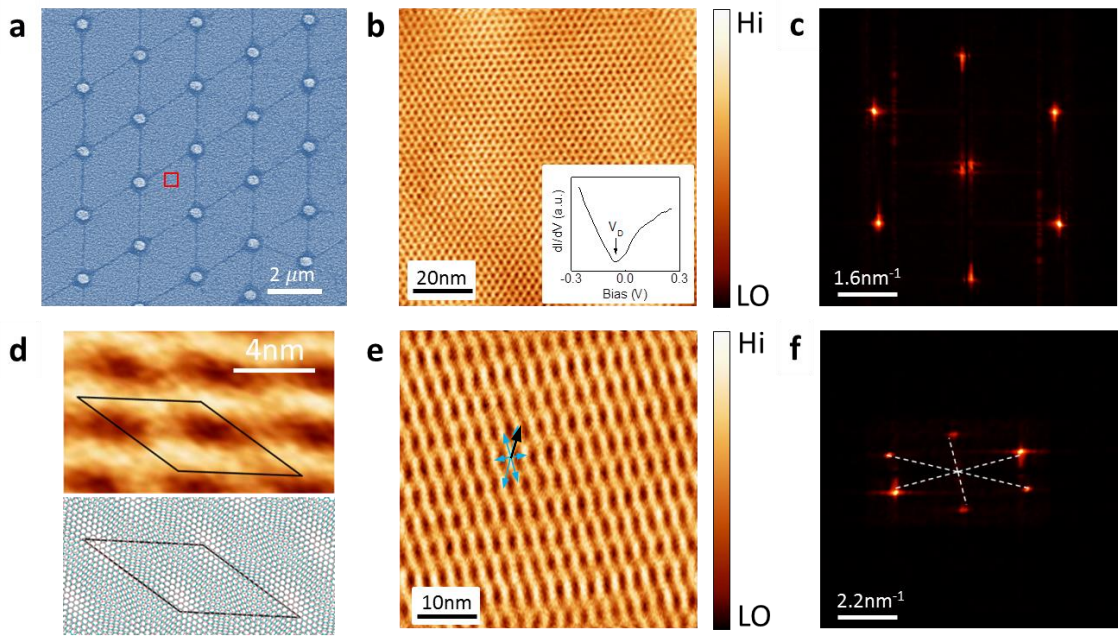

**Figure 3**

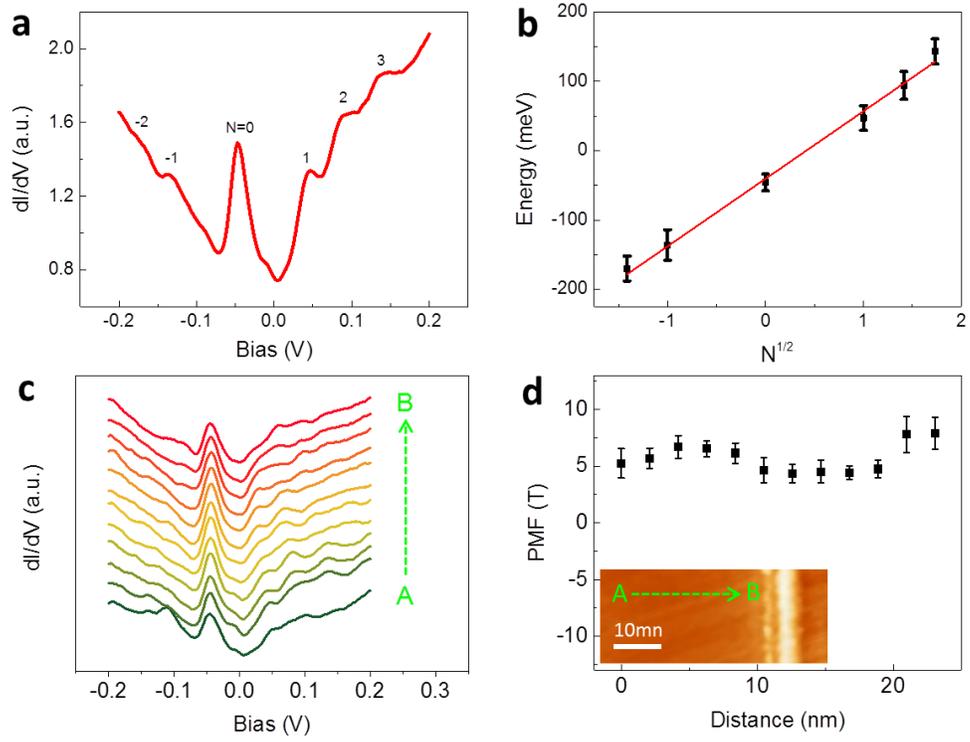

**Figure 4**

# Supplementary Information

## 1. Method for fabricating graphene on LOR pillars

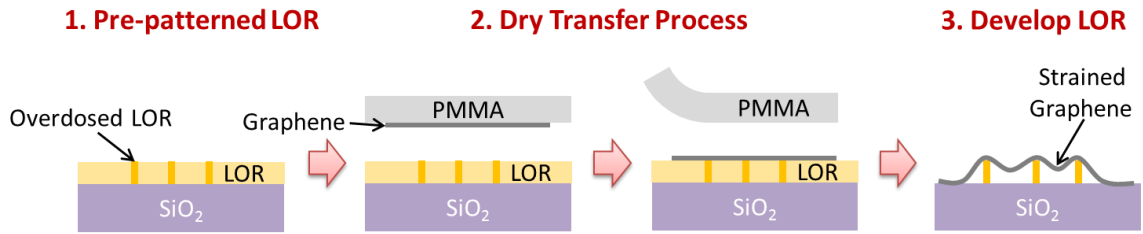

Figure 1s Schematic drawing of method for graphene on LOR pillars.

To fabricate the LOR pillar array we overdose the LOR with the electron beam but do not proceed to the development stage (Step 1 in Figure 1s), *i.e.*, the LOR pillars are still buried in the thin film, Figure 2a. Subsequently, a graphene flake on PMMA is transferred onto the undeveloped pillar array. To protect the graphene and LOR from exposure to acetone, the PMMA thin film is directly peeled off (Step 2 in Figure 1s). After the transfer, the whole sample is developed in ethyl-lactate to remove the LOR between the overdosed LOR pillars (Step 3 in Figure 1s). The sample is directly taken out from the liquid and dried with nitrogen gas. Due to the surface tension of the liquid, graphene between the pillars area will be dragged down. Inset of Figure 2b shows the AFM image of stretched graphene by LOR pillars.

## 2. Undistorted Moiré pattern between graphene and BN in unstrained graphene.

The distorted Moiré pattern between graphene and the hBN substrate, shown in Figure 3, was attributed to strain. However, in order to rule out artifacts such as could be produced by a faulty piezo-scanner, we directly check the Moiré structure between unstrained graphene and hBN. As shown in Figure2S. The super-pattern in the absence of strain is undistorted.

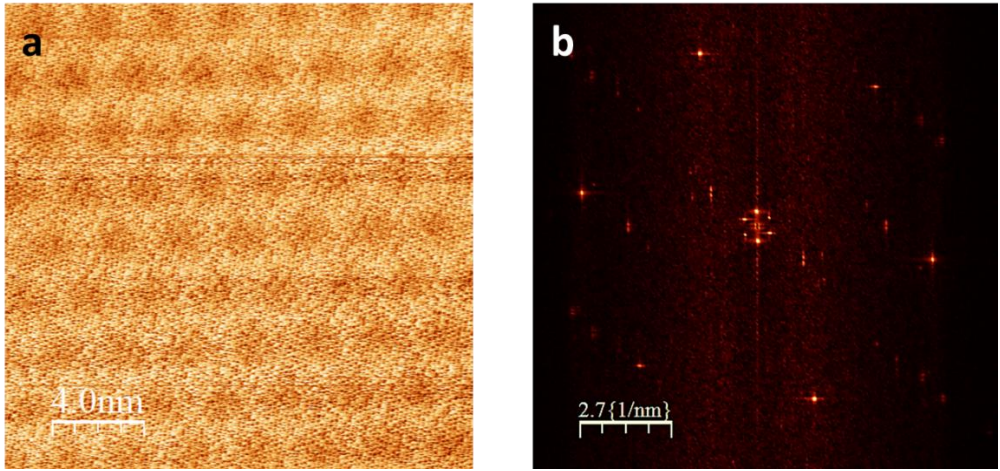

Figure 2s STM topography of an unstrained graphene flake deposited on an hBN substrate (a) and the corresponding FFT pattern (b). $V_b$=-300mV, I=20pA.

## 3. Modifying the strain in graphene by different pillars configuration.

In the main text, we showed the distorted moiré pattern between the strained graphene and the hBN substrate. Here, in order to illustrate the tunability of the induced strain in graphene by the nanopillars, we studied different G/Nanopillars/hBN samples by modifying the height of the pillars. Fig.3s (a) shows the topography of a different sample with 50nm height nanopillars. Clearly in this sample, the distorted moiré pattern is also observed. Using the model presented in the main text, we obtain $\varepsilon \sim 5\%$, and $\theta \sim 5°$ for this sample. These parameters are then used to simulate the distorted moiré pattern, Fig.3s (b), and compare to the measured pattern Fig.3s (a).

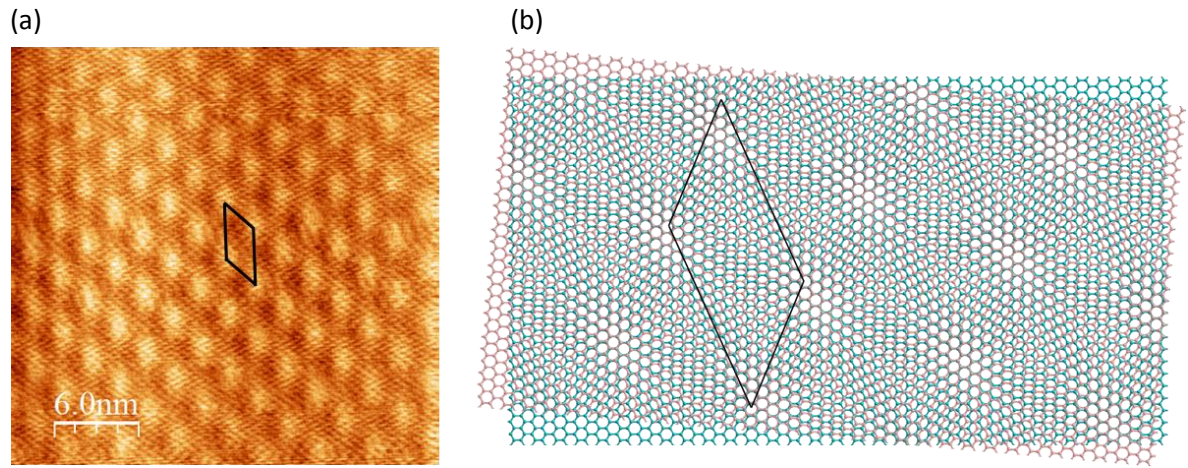

Fig.3s (a) STM topography of the distorted moiré pattern of stretched graphene on hBN. $V_b$=-300mV, I = 20pA. The black lines indicated the unit cell of the distorted moiré pattern. (b) Simulated distorted moiré pattern by superposing stretched graphene (red) on BN substrate (blue).

### 4. Magnifying glass effect for different twist-angles

Here we illustrate that the magnifying effect of the distorted moiré pattern is applicable to a large range of twist-angles. Fig.4s shows the simulated moiré pattern for different angles from 1° to 20°, which shares the same strain and Poisson ratio as the model in Fig.3. As one can see from the images, the magnifying effect persists up to angles at which the moire pattern period approaches the lattice constant.

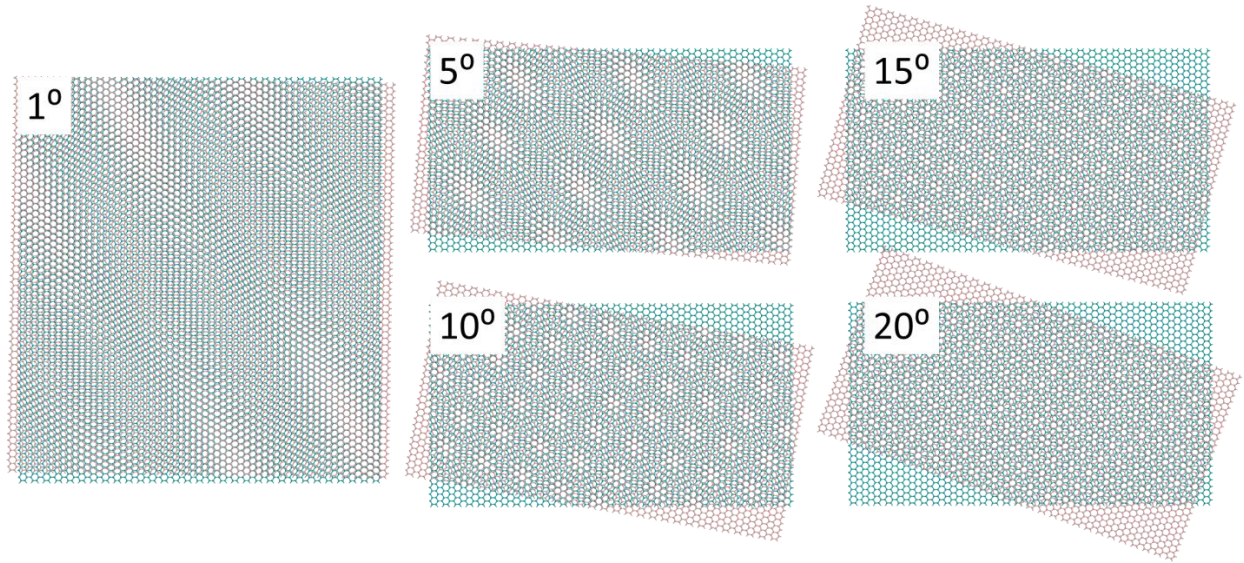

Fig.4s Evolution of the distorted moiré pattern with twist-angle. The blue lattice represents the hBN substrate and red lattice represents graphene. The numbers label the twist-angle.

**5. Backgate dependence of the dI/dV curves for the strained graphene.**

In order to demonstrate the robustness of the induced PMF, we show the backgate dependence of the dI/dV curves in Fig.5s.

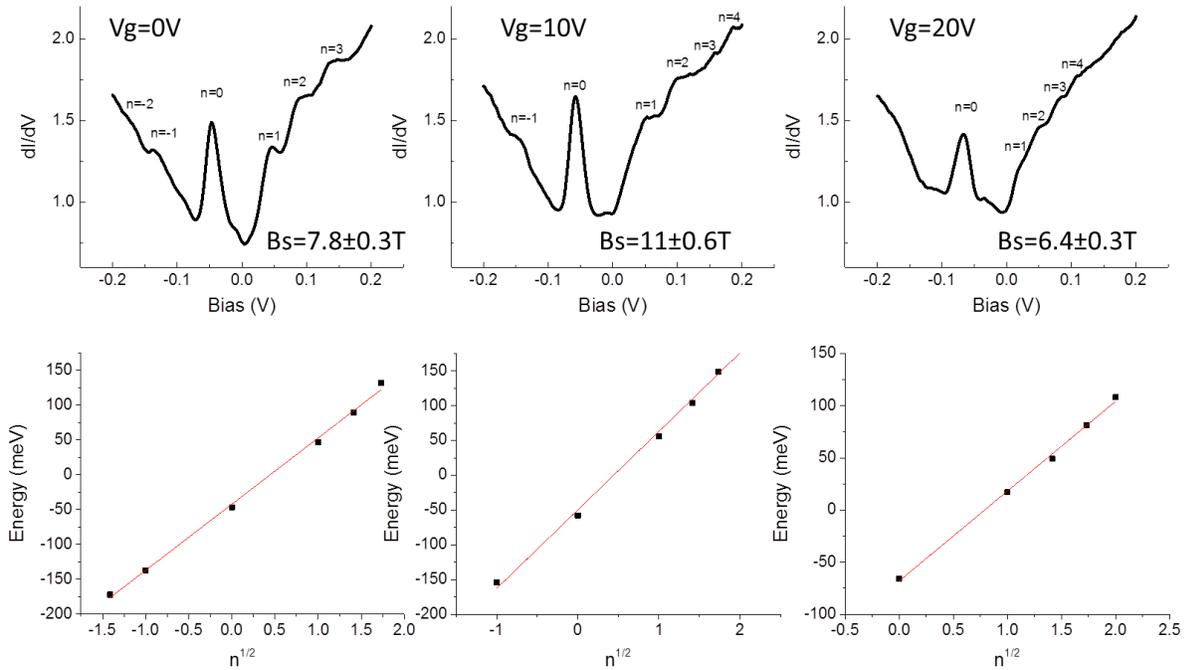

Fig.5s Backgate dependence dI/dV curves. The bottom panels show the fitting of the peaks' energy position versus the index which allows us to extract the pseudomagnetic value (main text see the details).

## 6. Analysis of the dI/dV curve to extract the Landau levels

In the main text, we showed the dI/dV curve on the strained graphene area, the peaks in the dI/dV curves are assigned to the PLLs, Fig.4a. Most peaks (N=-1, 0, 1, 2, 3) are easy to identify but the N=-2 peak is less pronounced. In order to simplify the identification process, the background is removed (a linear baseline was subtracted on the two sides of the N=0 peak), Fig. 6s.

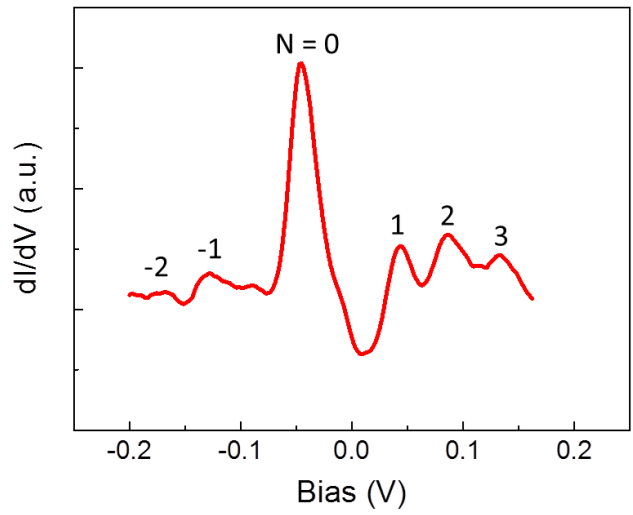

Fig.6s Same dI/dV curve as Fig.4a in the main text, but with the background removed.